\def\e{\begin{equation}}
\def\ee{\end{equation}}
\def\ea{\begin{eqnarray}}
\def\eea{\end{eqnarray}}
\def\nm{\nonumber \\ }

\def\v#1{\vert #1 \rangle}
\def\a{ab;cd}
\def\WA{$W\!\!A_2 $ algebra }
\def\a{\alpha}
\def\L{\Lambda}
\def\l#1{\lambda_{#1}}
\def\I{\vert I\vert }

\def\NPB#1#2#3{{\sl Nucl. Phys.} {\bf B#1} (#2) #3}
\def\PLB#1#2#3{{\sl Phys. Lett.} {\bf #1B} (#2) #3}
\def\FAP#1#2#3{{\sl Funkt. Anal. Prilozheniya} {\bf #1} (#2) #3}
\def\IJMPA#1#2#3{{\sl Int. J. Mod. Phys.} {\bf A#1} (#2) #3}

\documentstyle[12pt]{article}
\advance \topmargin by -\headheight
\advance \topmargin by -\headsep

\evensidemargin \oddsidemargin
\begin{document}

\title{  \begin{flushright}
\normalsize{ DAMTP  94-18 \\
                hep-th/9403032 }
  \end{flushright}
\vspace{5cm}
 Singular vectors of the $W\!\!A_2$ algebra }

\author{\large{ Z. Bajnok\thanks{ On leave from
 Institute for Theoretical Physics,  Roland E\"otv\"os   University,
 H-1088 Budapest, Puskin u. 5-7, Hungary}}
 \\ \\
 \normalsize{\it Department of Applied Mathematics and Theoretical Physics}\\
 \normalsize{\it University of Cambridge, Silver Street,   } \\
 \normalsize{ \it Cambridge, CB3 9EW, U.K.} }

\def\today{March 6, 1994 }

 \maketitle

\vspace{1.5cm}

 \begin{abstract}

The null vectors of an arbitrary highest weight representation of the
 $W\!\!A_2$ algebra are constructed. Using an extension of the
enveloping algebra by allowing complex powers of one of the
generators, analysed by Kent for the Virasoro theory, we generate
all the singular vectors indicated by the Kac determinant. We prove
that the singular vectors with  given  weights are unique up to
normalisation and consider the case when $W_0$ is not diagonalisable
among the singular vectors.

\end{abstract}

\newpage

The highest weight representations of the Virasoro algebra or its
extensions play an important role in the analysis of conformal
field theories. In fact, the degenerate representations
are crucial since they contain null vectors, which restrict the
possible fusions of the fields, give differential equations for
the correlation functions and determine the field content of
the theory. A careful analysis of their embedding pattern
makes  it possible to construct the characters of the irreducible
representations,
from which the modular invariant partition function of the model
 is built up. For this reason
knowledge of these null vectors and their embedding pattern is
essential.

Basically there are two  approaches to give explicit expressions for these
null vectors. Much of
the effort is based on one of these, the fusion procedure
 of  Bauer et al \cite{Bau}.
The other one uses complex powers of the generators and  was
introduced by Malikov, Feigin and Fuchs (MFF)
for the modules over KM algebras \cite{Mali}. Later Kent
extended  the method  for the Virasoro algebra \cite{Kent}.
 Ganchev and Petkova discovered a relationship between them by
considering
the reduction of the MFF singular vectors \cite{Ga}. Analysing the
reduction of the $\hat {sl_3}$ singular vectors they
proposed a similar extension  for the $W\!\!A_2$ algebra  \cite{Fur}.

The aim of the paper is to set up consistently this extended  space, i.e.
following the earlier works we generalise the method of
complex powers for the $W\!\!A_2$ algebra. Contrary to the naive hope,
instead of using complex powers of the mode $L_{-1}$  we consider
complex powers of the generator $W_{-1}$. Using this extension
which allows us to define an analytic continuation of the null
vectors, found explicitly by Bowcock and Watts \cite{BoWa},
we generate singular
vectors in arbitrary h.w.\  representations.
In particular we show that  the h.w.\  singular vectors with given weights
are unique up to normalisation.  We construct
singular vectors that are not eigenstates of the $W_0$, but are
annihilated by the positive modes.  Including these non h.w.\  type
singular vectors we produce all the singular vectors given by the
Kac formula \cite{Wa}.

First we recall the results of the degenerate
representations of the algebra.
The \WA   is defined by the following  commutation relations:
\ea
  [L_{n},L_{m}] &=&(n-m)L_{n+m}+{c\over 12}n(n^2-1)\delta _{n+m}  \nm
 {}~[ L_{n},W_{m} ] &=&(2n-m)W_{n+m} \nm
 {}~[ W_{n},W_{m} ] &=&{{22+5c}\over 48}{c\over 3\cdot 5!}(n^2-4)(n^2-1)
   n\delta_{n+m}\nm && +{1\over 3}(n-m)\Lambda _{n+m}
+{22+5c\over 48}{1\over 30}(n-m)(2n^2-nm+2m^2-8)
L_{n+m}
\eea
here the  $\Lambda$ field is related to the
$ (L_{-2}L_{-2}-{3\over 5}L_{-4}) \v{0} $
state in the sense of the meromorphic conformal field theory  \cite{God},
 and we used the normalisation of \cite
{BoWa}.

A highest weight representation is characterised by its
$h,\ w, $ and $c$ values and contains a h.w.\  state which satisfies:
\e
 L_n\v{hw}=\delta_{n,0} h \v{hw} \quad ; \quad  W_n\v{hw}
 =\delta_{n,0}w \v{hw} \quad ; \quad  n\geq 0
\ee
The Verma module, for which we write  $V(h,w,c)$,
is generated by acting with the negative
modes $L_{-n},\ W_{-m} $ on the h.w.\  state.
(Sometimes we write $W^2 $ for $L$ and $W^3$ for $W$).
A basis of the space is defined by using the following ordering:
\e
W_{-I}\v{hw}=W_{-i_1}^{j_1} W_{-i_2}^{j_2}\dots  W_{-i_n}^{j_n}\v{hw}
= W_{-i_1}\dots  W_{-i_k}L_{-i_{k+1}}\dots L_{-i_n}\v{hw}
\ee
where $i_l>0$, $ j_l \ge j_{l+1}$, and if $j_l=j_{l+1}$ for some $l$ then
$i_l \le i_{l+1}$. Here $I$ denotes the collection of the $j_l$ and $i_l$
numbers ($I=\{i_1,\dots,i_{k};i_{k+1},\dots,i_n\}$)
 and $\vert I \vert =\sum_{l=1}^n i_l $ is the level of the
corresponding state. Using the
natural grading on this representation space given by the
eigenvalues of $L_0$ we can write:
\e
V(h,w,c)=\bigoplus_{0,1,2 \dots }V_n(h,w,c)
\ee
where $V_n(h,w,c)$ is the eigenspace of $L_0$ with eigenvalue $h+n$.

In order to analyse the reducibility of the representation space
we introduce the following parametrisation of the $W$ weights:
\e
h={1\over 3}(x^2+xy+y^2)-(\a-{1\over \a})^2
 \qquad w={1\over 27}(x-y)(2x+y)(x+2y)
\ee
where $c=2-24( \a-{1\over \a})^2 $.
(Sometimes we write $t$ for ${1\over {\a ^2}}$).
We associate with each h.w.\ representation of the \WA  a h.w.\
representation of
 $sl_3$ described by the $\L (x,y)=x\l{1}+
y\l{2} $ weight. Here the $\l{i}$s are the two fundamental weights
 of $sl_3$.
We remark that the parametrisation of the $W$ weights is
redundant i.e. the $\Lambda $ weight is well defined up to a
Weyl reflection. In other words not only $\Lambda (x,y)$ but also
 $\Lambda (-x,x+y)$, $\Lambda (-y,x+y) $,
 $\Lambda (-x-y,x)$, $\Lambda (y,-x-y)$ and $\Lambda (-y,-x)$
correspond to the same $h$ and $w$.

We know from the determinant formula that if $x=a\a-{c\over \a}$ or
$y=b\a-{d\over \a}$, with positive integers $(a, c)$ or $(b, d)$,
 then the representation of the $W$ algebra
 is reducible (degenerate) and there is  a h.w.\
 null state at level $ac$ or $bd$. If both hold the representation is
called doubly degenerate. We  call a vector singular or null   if it
is annihilated by the positive modes.
A null vector is not always a  h.w.\ vector since sometimes $W_0$ is not
diagonalisable.
Clearly null vectors define invariant subspaces. In the case of
degenerate representation we denote the h.w.\  vector of the Verma module
by $(ab,cd)$ or $\v{ab,cd}$.

Now we summarise the well-known results of the representation theory.
In the case of degenerate representation (say $a,c $ are positive
  integers)
there is a null state at level $ac$.
This is a h.w.\ state, which  can be parametrised by $(-a,a+b;cd)$, i.e.
 it corresponds to the weight
$\L^{'}=\L-a\a e_1$ of $sl_3$, where $e_1$ is its first simple
root, and  is an eigenvector of $W_0$ with the appropriate weight.
Unfortunately the general form of the operator, which generates
the null state from the h.w.\  state, is not known in terms of the
natural variables of the $W$ algebra. We have the following partial result:
in the $(pq,11)$ representation  the following operator generates a singular
vector at level $p$:
\e
{\cal O}_{p,1} =\sum_{\{n_i\}: \sum_{i=1}^r n_i=p}
 c_{\{n_i\} }(p,q,\a)\prod_{i=1}^r \biggl (
W_{-n_i}-\a[{1\over 6}(n_i(
t-3)+2(t-p-2q))+N_i]L_{-n_i} \biggr )
  \label{O}
\ee
Where
\e
 c_{\{n_i\} }(p,q,\a)=(-\a)^{p-r}{\prod_{  {
         i=1} \atop {i\neq N_k ;k=1,\dots r-1}}^{p-1}
i(i-p)(i+t-p-q)\a^2}
\quad ; \quad
 N_k=\sum_{j=1}^k n_j
\ee
Furthermore ${\cal O}_{p,1}\v{pq,11}$ defines a h.w.\  singular vector with
weight $(-p,p+q;11)$ \cite{BoWa,Fur}.
The operators generating null vectors at level $p$ in the representation
$ \v{qp,11}$, $\v{11,pq}$, $\v{11,qp}$ can be obtained from
(\ref{O}) by $\a \to -\a$, $ \a \to -{1\over \a}$ and
 $ \a \to {1\over \a}$,
respectively.

 Generalising the method of ref. \cite{Mali,Kent}
we extend the enveloping algebra by allowing complex powers of
the generator $W_{-1}$. The commutation relations with the new generator
are:
\e
[(W_{-1})^a,W_n^k]=\sum_{j=1}^{\infty}{a \choose j}
   \biggl [W_{-1}, ... \Bigl [W_{-1},  \bigl [ W_{-1},W_n^k
     \bigr ]  \Bigr ] ... \biggr ](W_{-1})^{a-j}
\ee
or
\e
[W^k_n,(W_{-1})^a]=\sum_{j=1}^{\infty}{a \choose j}(W_{-1})^{a-j}
   \biggl [ ... \Bigl [ \bigl [ W_n^k,W_{-1} \bigr ],W_{-1}
    \Bigr ] ...W_{-1} \biggr ]
\ee
where $k=2,3$ and
 ${a \choose j}={a(a-1)\dots (a-j+1)\over {j!}}$. In both cases
we have the commutator of $W_n^k$ with $W_{-1}$ $j$ times.
This formula is valid not only for $W_n^k$ but for arbitrary
$W_{-I}$.

We define the generalised Verma module, ($\tilde V(h,w,c)$),
 that can be obtained
by acting with  $(W_{-1})^a$ on the elements of the
 original Verma module, for arbitrary complex number $a$.
 We use the same ordering as before
but now allowing complex powers of $W_{-1}$.
Due to the grading given by $L_0$ we can split the representation
space into eigenspaces of this operator. We call an operator
well defined if it has the following form:
\e
{a_0(W_{-1})^a + \sum_I a_I  (W_{-1})^{a-\I}}W_{-I}
\ee
Here $W_{-I}$ contains negative modes only and
 we allow the sum to be infinite, but all the coefficients have to
be finite. From now on we use a new terminology: we call the term
$a_I  (W_{-1})^{a-\I}W_{-I}$
in the above sum an operator at level $\I$.

 This framework we have just established
is a consistent framework in the sense that if
any well defined operator acts on a state of the generalised Verma
module, which was created from the h.w.\  state
 by some other well defined operator, then
the resulting state is also  created by some well defined operator.
To see this
one has to consider the product of two well defined operators
acting on $  \v{hw} $
and reorder it into the usual basis.
\ea
 & \Bigl (a_0(W_{-1})^a + \sum_I a_I  (W_{-1})^{a-\I}W_{-I}\Bigr )
\Bigl (b_0(W_{-1})^b + \sum_{I'} b_{I'}(W_{-1})^{b-\I'}W_{-I'}\Bigr )
 \v{hw}
\nm & = \Bigl( c_0(W_{-1})^c + \sum_{I''} c_{I''} (W_{-1})^{c-\I''}W_{-I''}
 \Bigr ) \v{hw} \quad ,
\eea
where  $c=a+b$. Clearly the only thing we have to show is that all
the $c_{I''}$ coefficients are finite.  Considering the commutation
relations we can see that the operators
which contribute at level $\vert I''\vert $
 have level less than or equal to ${4\over3} \vert I'' \vert$.
 The number of these levels are finite and  furthermore
each level gives only a finite contribution for the level in
question so this shows that the contributions are finite.
(In order to  see the factor
 ${4\over3}$ we remark that we can obtain terms, which contain
$W_{-1}$ from the reordering of $W_{-I}$ and $W_{-I'}$. However from the
commutation relation it follows that we need three $W$ modes to produce one
$W_{-1}$. Considering in a
 similar manner the nested commutators of $W_{-I}$
with $W_{-1}$
we can say that the number of $W_{-1}$ obtained in this way is at most one
quarter of the level,
say for example ($[[W_{-2},W_{-1}],W_{-1}] \sim W_{-1}$).)

For later applications we have to show that all singular vectors
at level zero are proportional to $\v{hw}$. The proof is rather lengthy
so we just sketch it here.
First we suppose that this is not the case
i.e.
\e
\sum _{I,\I\neq 0} a_I ( W_{-1})^{-\I}W_{-I}\v{hw}
\ee
 is a h.w.\  singular vector, with $W_0$ eigenvalue $w$.
Then we define an ordering among the modes, we say that
$ I=\{i_1,\dots ,i_k;i_{k+1},\dots,i_n\}$ precedes
$ J=\{j_1,\dots ,j_l;j_{l+1},\dots,j_m\}$ if $\I <  \vert J\vert $ or if
$\I = \vert J\vert $ then there exists an $r$ such that $i_r<j_r$.
Considering the commutations relations we can see that acting on $W_{-I}$
with an $L$ ($W$) mode we can get at most  $[{k+2\over3}]$
($[{k\over3}]+1$) $W_{-1}$s, respectively,
where $[x]$ denotes the integer part of $x$.
The idea of the proof is that we suppose that
  $ a_I =0$ if $\I<k$ or if $W_{-I}$ can
give contributions to a  smaller level then $k-1$. From this hypothesis
we shall show that the coefficients at level $k$ and the coefficients
 which can contribute at level $k-1$ have to vanish.
However these terms at level $k+l$ contain precisely
$3l, 3l+1$ or $3l+2$ $W$ modes. Since the  coefficients
which contain more $W$s are zero due to the induction hypothesis
 it is not hard to see that all the coefficients have to vanish.
First we consider the highest level which contribute and has non
zero terms. Then we take the term which contains the most $W$ and
the smallest in the ordering defined above
with  $ I=\{i_1,\dots ,i_k;i_{k+1},\dots,i_n\}$. Acting on it
with $ L_{i_1-1} $ (   $ W_{i_1-1}$ if $k=0$) we find that
the coefficient of $ (W_{-1})^{-\I+1}W_{-I'}$ is not zero. (Here we
obtained $I'$ deleting $i_1$ from $I$).
Since this was the only term which contributes we conclude
that  $a_I$ vanishes. Following the same procedure from level to level
we can show that all the coefficients have to vanish. The only non
trivial thing is that if we consider a term at level $k+l$ which contains
 $3l$ $W$ modes then we can get contributions from the terms which
contain  $3l-1$ $W$ modes at level $k+l-1$. However taking under
investigation these terms first then later  the other this problem
can be solved easily.
We remark that $I$ may contain the mode $L_{-1}$ so we really have to
use the fact that the vector is an eigenvector of $W_0$.
 Unfortunately this proof can not be applied to the case when the algebra
is extended with complex powers of $L_{-1}$, although the commutation
relations are much more simple in this case.

Now we are ready to give a generalisation of the null vectors (\ref{O}).
First we reorder the expression into our basis:
\e
{\cal O}_{p,1}\v{pq,11}= \Bigl((W_{-1})^p +\sum_{I,\I \le p}a_I(p,q,\a)
                          ( W_{-1})^{p-\I} W_{-I}\Bigr)\v{pq,11}
\ee
 Then we show that the $a_I(p,q,\a)$  coefficients are polynomials
in $p$ for fixed $I$. Since $a_I(p,q,\a)$ contains
only sums of
products of polynomials of $p$, what we need is that these products
are still polynomials in $p$.
 Considering the original expression, (\ref{O}),
only those terms can give contributions to  $a_I(p,q,\a)$
 whose $W_{-1}$s  are between
$( W_{-1})^p$ and $( W_{-1})^{p-{4\over3}\I}$.  They
correspond to the partitions which have at least $p-{4\over3}\I$ ones.
However
the number of these partitions can be estimated as a polynomial in $p$
with a power which is a function of $\I$. If we study the contributions of each
partition we see that the product in (7) and the coefficient of the
$L$ modes  contain less factors than
${4\over3}\I$.
 Moreover if we observe that the number of
commutations are functions of $\I$, collecting all the powers,
 we can conclude that the
coefficients after the ordering are polynomials in $p$.
However this means that we can make a unique analytic continuation
of this operator and so we have:
\e
{\cal O}_{a,1}\v{ab,11}= \Bigl ((W_{-1})^a +\sum_{I}a_I(a,b,\a)
                            ( W_{-1})^{a-\I} W_{-I}\Bigr )\v{ab,11}
\ee
Here $a$ and $b$ are arbitrary complex numbers and the operator
which creates this state is a well defined operator.
 In the notation we suppress the dependence on $q$ and $\a$ if it is clear
from context.

Now we show that this vector is a singular vector in the  $(ab,11)$
representation at level $a$. Let us act first with the operator
$L_1$:
\e
L_1{\cal O}_{a,1}\v{ab,11}=\sum_{I}Q_I(a,b,\a)
                             ( W_{-1})^{a-1-\I} W_{-I}\v{ab,11}
\ee
Here $Q_I(a,b,\a)$ is a linear combination of
$a_{I'}(a,b,\a)$, where $\vert I' \vert \leq {4\over 3}\I+2$
and this way clearly is a polynomial in $a$. If we take $a$ to be
an integer larger than  ${4\over 3}\I+2$ then  $Q_I(N,b,\a)$ is the coefficient
of an operator which is an element of the enveloping algebra and
which we would find in $L_1{\cal O}_{N,1}\v{Nb,11}$. Since this coefficient
really has to vanish we conclude that $Q_I(a,b,\a)$ is a polynomial
in $a$ which  vanishes for every integer larger than  ${4\over 3}\I+2$,
so it is identically zero.
Similarly we can prove that this vector is a h.w.\  vector, i.e.\
it is annihilated by the positive modes and can be described by
the $(-a,a+b)$ h.w.\  state.

{}From now on we write $(a,b)$ for the $(ab,11)$ h.w.\  vector.
The remnants of the parametrisation ambiguity mentioned earlier
 in the language of $(a,b)$ are
the shifted Weyl reflections i.e. the
equivalent
parametrisations are: $ (b,-a-b+3t) $, $ ( -b+2t,-a+2t)$,
$(-a-b+3t,a)$, $(a+b-t,-b+2t)$ and $ (-a+2t, a+b+t)$.
 This means that there are at least two ways to generate
null vectors in a certain representation.

Now we shall show that we have only one singular vector at a certain level
with a given $W_0$ eigenvalue. To see this first we remark that the
null vectors, which are generated by some ${\cal O}_{a,1}$, have
always $(W_{-1})^a$ leading coefficient. Suppose that we generated
a singular vector at level $a$: ${\cal O}_{a,1}\v{ab}$.
Since this h.w.\  state can be described by  the $(-a,a+b)$ parameters,
we have a singular vector at level zero, namely:
\e
{\cal O}_{-a,1}{\cal O}_{a,1}\v{ab,11}=(1+(W_{-1})^{-1}(\dots))
               \v{ab,11}= \v{ab,11}
\ee
In the second equality we used the fact that the only singular
vector at level zero is the h.w.\  state itself.
If there is another singular vector at this level with
the same $W_0$ eigenvalue then it may have leading coefficient
 $(W_{-1})^a$. Since the null vectors are not the same, as we supposed,
taking their difference we have a singular vector at level $a$
 with leading coefficient at most $(W_{-1})^{a-1}$. But acting now with
${\cal O}_{-a,1}$ on this h.w.\  state we obtain a singular vector at
level zero, which does not contain $\v{ab,11}$, i.e. it is zero.
However this implies that the vector is zero itself.
Summarising we found that the only singular vector at level $a$
with weight $(-a,a+b)$ in the $(a,b)$ representation
is the same as that which we can generate, and so  it is unique.

 Now we take a representation described by $(a,b)$.
Acting on it with one of the operators, say with ${\cal O}_{b1}$,
the representation obtained can be described by $(a+b,-b) $. Clearly
this weight is connected to the original h.w.\  by a Weyl reflection.
We can use now the shifted Weyl reflections and reparametrise it as
 $(a-t,b+2t) $. So we can act again and obtain
the $(a-2t,b+4t) $ representation. Using the six directions, given by
the reparametrisation invariance, acting with the appropriate
operators  we obtain the following picture, which looks like the weight
diagram of $sl_3$:

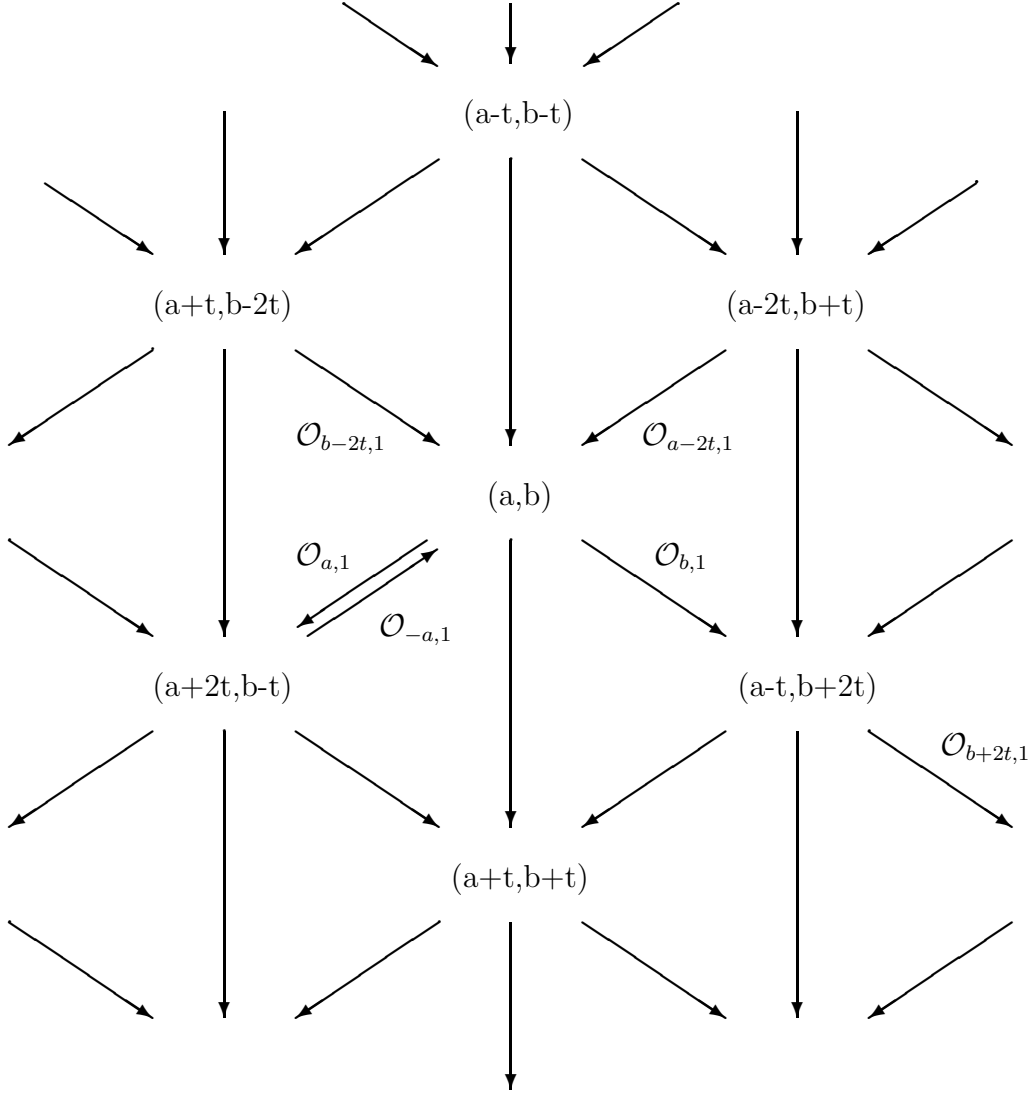
\setlength{\unitlength}{0.0125in}%
\begin{picture}(190,500)(110,305)
\thicklines
\put(380,580){\vector( 3,-2){ 60}}
\put(440,500){\vector(-3,-2){ 60}}
\put(260,500){\vector( 3,-2){ 60}}
\put(350,580){\makebox(0.4444,0.6667){\tenrm .}}
\put(350,580){\vector( 0,-1){120}}
\put(440,660){\vector(-3,-2){ 60}}
\put(260,660){\vector( 3,-2){ 60}}
\put(380,420){\vector( 3,-2){ 60}}
\put(320,420){\makebox(0.4444,0.6667){\tenrm .}}
\put(320,420){\vector(-3,-2){ 60}}
\put(500,500){\makebox(0.4444,0.6667){\tenrm .}}
\put(500,500){\vector( 3,-2){ 60}}
\put(560,580){\vector(-3,-2){ 60}}
\put(500,660){\vector( 3,-2){ 60}}
\put(200,660){\makebox(0.4444,0.6667){\tenrm .}}
\put(200,660){\vector(-3,-2){ 60}}
\put(140,580){\makebox(0.4444,0.6667){\tenrm .}}
\put(140,580){\vector( 3,-2){ 60}}
\put(200,500){\vector(-3,-2){ 60}}
\put(140,420){\makebox(0.4444,0.6667){\tenrm .}}
\put(140,420){\vector( 3,-2){ 60}}
\put(230,500){\makebox(0.4444,0.6667){\tenrm .}}
\put(230,500){\vector( 0,-1){120}}
\put(230,660){\vector( 0,-1){120}}
\put(470,660){\vector( 0,-1){120}}
\put(470,500){\vector( 0,-1){120}}
\put(560,420){\vector(-3,-2){ 60}}
\put(350,420){\vector( 0,-1){ 70}}
\put(320,740){\vector(-3,-2){ 60}}
\put(380,740){\vector( 3,-2){ 60}}
\put(155,730){\vector( 3,-2){ 45}}
\put(230,760){\vector( 0,-1){ 60}}
\put(470,760){\vector( 0,-1){ 60}}
\put(350,740){\makebox(0.4444,0.6667){\tenrm .}}
\put(350,740){\vector( 0,-1){120}}
\put(350,805){\vector( 0,-1){ 25}}
\put(545,730){\makebox(0.4444,0.6667){\tenrm .}}
\put(545,730){\makebox(0.4444,0.6667){\tenrm .}}
\put(545,730){\vector(-3,-2){ 45}}
\put(280,805){\makebox(0.4444,0.6667){\tenrm .}}
\put(420,805){\makebox(0.4444,0.6667){\tenrm .}}
\put(280,805){\vector( 3,-2){ 39.231}}
\put(420,805){\vector(-3,-2){ 39.231}}
\put(315,580){\vector(-3,-2){ 54.231}}
\put(265,540){\vector( 3, 2){ 54.231}}
\put(340,595){\makebox(0,0)[lb]{\raisebox{0pt}[0pt][0pt]{\twlrm (a,b)}}}
\put(330,755){\makebox(0,0)[lb]{\raisebox{0pt}[0pt][0pt]{\twlrm (a-t,b-t)}}}
\put(325,435){\makebox(0,0)[lb]{\raisebox{0pt}[0pt][0pt]{\twlrm (a+t,b+t)}}}
\put(200,675){\makebox(0,0)[lb]{\raisebox{0pt}[0pt][0pt]{\twlrm (a+t,b-2t)}}}
\put(440,675){\makebox(0,0)[lb]{\raisebox{0pt}[0pt][0pt]{\twlrm (a-2t,b+t)}}}
\put(200,515){\makebox(0,0)[lb]{\raisebox{0pt}[0pt][0pt]{\twlrm (a+2t,b-t)}}}
\put(445,515){\makebox(0,0)[lb]{\raisebox{0pt}[0pt][0pt]{\twlrm (a-t,b+2t)}}}
\put(260,620){\makebox(0,0)[lb]{\raisebox{0pt}[0pt][0pt]{${\cal O}_{b-2t,1}$}}}
\put(405,620){\makebox(0,0)[lb]{\raisebox{0pt}[0pt][0pt]{${\cal O}_{a-2t,1}$}}}
\put(360,675){\makebox(0,0)[lb]{\raisebox{0pt}[0pt][0pt]{}}}
\put(260,570){\makebox(0,0)[lb]{\raisebox{0pt}[0pt][0pt]{${\cal O}_{a,1}$}}}
\put(410,570){\makebox(0,0)[lb]{\raisebox{0pt}[0pt][0pt]{${\cal O}_{b,1}$}}}
\put(530,490){\makebox(0,0)[lb]{\raisebox{0pt}[0pt][0pt]{${\cal O}_{b+2t,1}$}}}
\put(355,520){\makebox(0,0)[lb]{\raisebox{0pt}[0pt][0pt]{}}}
\put(260,305){\makebox(0,0)[lb]{\raisebox{0pt}[0pt][0pt]{ }}}
\put(295,540){\makebox(0,0)[lb]{\raisebox{0pt}[0pt][0pt]{${\cal O}_{-a,1}$}}}
\end{picture}

Each arrow indicates that we have an operator
that generates a singular vector in the representation.
If we have a singular vector in $(a,b)$ at level $a$ then
we have a singular vector in $(a+2t,b-t)$ at level $-a$.
This shows that if there is an arrow which goes in one way then there
is another one which goes in the opposite way.
The diagram is commutative since the singular vectors generated
are unique.

We are now ready to generate null vectors in the generic case,
i.e. in the $(rp,sq)$ representation space. First we remark
that this representation is equivalent to the $(ab,11)$
representation if $a=r-(s-1)t$ and $b=p-(q-1)t $.
The null state in this representation has  parameters $(-r,r+p,sq)$
 or equivalently parameters  $(a^{'}b^{'},11)$ where
$a^{'}=-r-(s-1)t $ and $b^{'}=r+p-(q-1)t $. Of course we can use
the reparametrisation and  write $a{'}=r+(s+1)t $ and
$b{'}= p-(s+q-1)t $ for this state. However this state can be
obtained in the following form:
\e
{\cal O}_{r,s}\v{rp,sq}=
{\cal O}_{r+(s-1)t,1}{\cal O}_{r+(s-3)t,1}\dots
 {\cal O}_{r-(s-3)t,1} {\cal O}_{r-(s-1)t,1}\v{rp,sq}
\ee
Since this singular vector has the same weight as that which is
given by the Kac formula
 and the singular vectors are unique we conclude that the
state (17) is an element of the Verma module and defines
a null state at level $rs$.

  We remark that there are a lot
of other ways to generate the same singular vector.
We can reparametrise the $\v{rp,sq}$ h.w.\  vector  as $\v{11,ab}$ and
use the other type of operator ${\cal O}_{1,a}$, which have
 similar properties as the  ${\cal O}_{a,1}$ operators. Since
they do not give any new information about the theory and each singular
vector is unique (so it is enough for us to
 generate it one way) we are not
concerned with the other type of operators.

 In order to obtain the embedding relations of the singular
vectors  we have to consider
 those h.w.\ states of the previous diagram which
  are at positive integer level.
This condition
  can be formulated as follows:
after $m$ steps right  and $n$ steps left  the level is  a positive
 integer, i.e.
$np +mr-t(n(q-n)-m(s+n-m))$ is a positive integer.
We have to handle in a different way the case when $t$ is irrational or
rational.

First we consider the irrational case. Clearly in this case the coefficient
of $t$ has to be zero. This means that we have the following
possibilities: $n=q,m=0$;  $n=0,m=s$;  $n=q+s,m=s$;  $n=q,m=q+s$
or   $n=q+s,m=q+s$. However these null vectors are exactly the same
we would obtain from the character formula \cite{Kac} or from the
Kac determinant and the  embedding picture
looks like the Weyl reflections of the top h.w.\ vector.
\bigskip
 \bigskip

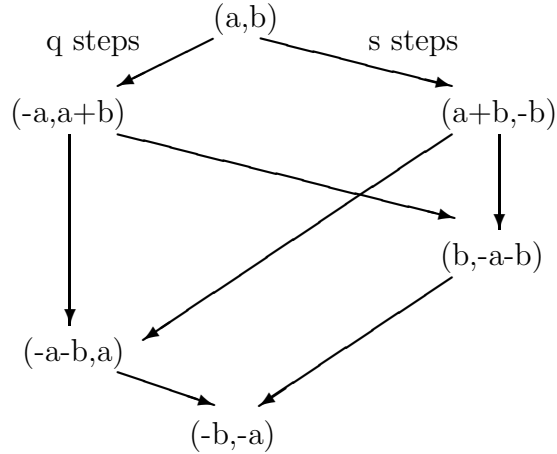
\setlength{\unitlength}{0.0125in}%
\begin{picture}(85,190)(135,470)
\thicklines
\put(340,640){\vector(-2,-1){ 40}}
\put(360,640){\vector( 4,-1){ 80}}
\put(300,600){\vector( 4,-1){141.177}}
\put(280,600){\vector( 0,-1){ 80}}
\put(460,600){\vector( 0,-1){ 40}}
\put(440,540){\vector(-3,-2){ 80.769}}
\put(300,500){\vector( 3,-1){ 40.500}}
\put(440,600){\vector(-3,-2){129.231}}
\put(340,645){\makebox(0,0)[lb]{\raisebox{0pt}[0pt][0pt]{\twlrm (a,b)}}}
\put(255,605){\makebox(0,0)[lb]{\raisebox{0pt}[0pt][0pt]{\twlrm  (-a,a+b)}}}
\put(435,605){\makebox(0,0)[lb]{\raisebox{0pt}[0pt][0pt]{\twlrm (a+b,-b)}}}
\put(260,505){\makebox(0,0)[lb]{\raisebox{0pt}[0pt][0pt]{\twlrm (-a-b,a)}}}
\put(435,545){\makebox(0,0)[lb]{\raisebox{0pt}[0pt][0pt]{\twlrm (b,-a-b)}}}
\put(330,470){\makebox(0,0)[lb]{\raisebox{0pt}[0pt][0pt]{\twlrm (-b,-a)}}}
\put(270,635){\makebox(0,0)[lb]{\raisebox{0pt}[0pt][0pt]{\twlrm  q steps}}}
\put(405,635){\makebox(0,0)[lb]{\raisebox{0pt}[0pt][0pt]{\twlrm s steps}}}
\end{picture}

\vskip 1cm

Now we consider the case of rational $t$, say $t={u\over v}$.
Furthermore we suppose that $p+q<r+s <u<v$. They correspond to
the minimal models. It is not hard to see
that moving $v$ units in each direction as many times as we want
we arrive always at a positive
 integer level. Moreover we will have the same
embedding picture beginning with each  point as we had in the
irrational case. This means that in
the following diagram the hexagons remind us of the earlier embedding
diagram with size $q$ and $s$ and they build a $v$ periodic lattice:

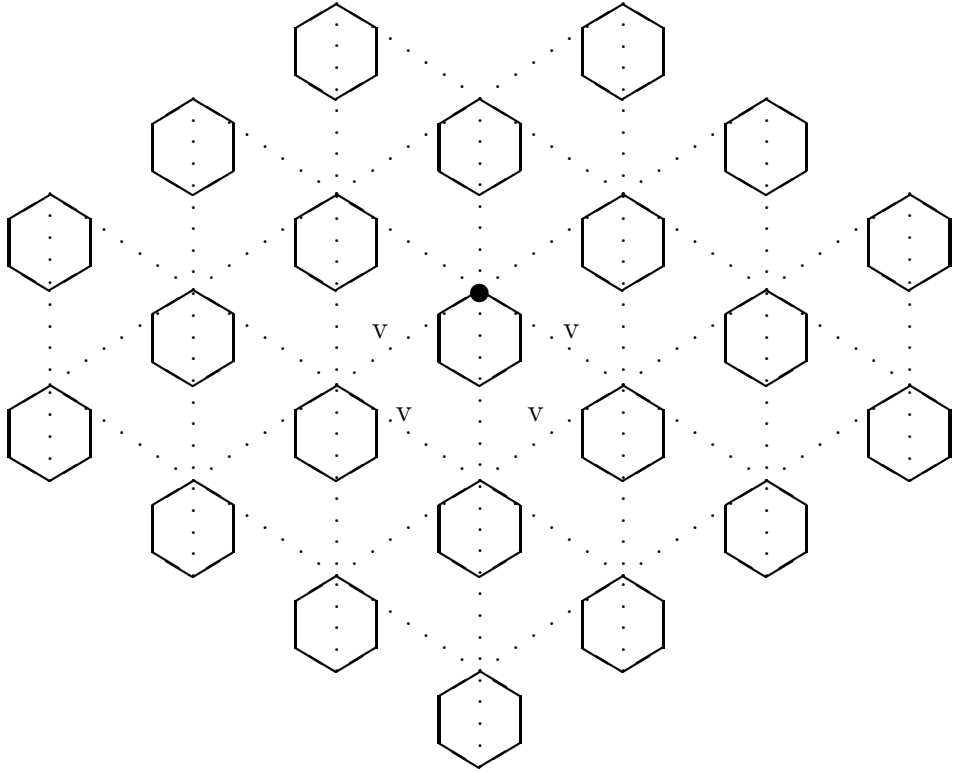
\setlength{\unitlength}{0.0125in}%
\begin{picture}(394,355)(150,365)
\thicklines
\put(400,599){\circle*{8}}
\put(400,440){\line(-5,-3){ 16.912}}
\put(383,430){\line( 0,-1){ 20}}
\put(383,410){\line( 5,-3){ 16.912}}
\put(400,400){\line( 5, 3){ 16.912}}
\put(417,410){\line( 0, 1){ 20}}
\put(417,430){\line(-5, 3){ 16.912}}
\put(460,560){\line(-5,-3){ 16.912}}
\put(443,550){\line( 0,-1){ 20}}
\put(443,530){\line( 5,-3){ 16.912}}
\put(460,520){\line( 5, 3){ 16.912}}
\put(477,530){\line( 0, 1){ 20}}
\put(477,550){\line(-5, 3){ 16.912}}
\put(460,480){\line(-5,-3){ 16.912}}
\put(443,470){\line( 0,-1){ 20}}
\put(443,450){\line( 5,-3){ 16.912}}
\put(460,440){\line( 5, 3){ 16.912}}
\put(477,450){\line( 0, 1){ 20}}
\put(477,470){\line(-5, 3){ 16.912}}
\put(520,520){\line(-5,-3){ 16.912}}
\put(503,510){\line( 0,-1){ 20}}
\put(503,490){\line( 5,-3){ 16.912}}
\put(520,480){\line( 5, 3){ 16.912}}
\put(537,490){\line( 0, 1){ 20}}
\put(537,510){\line(-5, 3){ 16.912}}
\put(520,600){\line(-5,-3){ 16.912}}
\put(503,590){\line( 0,-1){ 20}}
\put(503,570){\line( 5,-3){ 16.912}}
\put(520,560){\line( 5, 3){ 16.912}}
\put(537,570){\line( 0, 1){ 20}}
\put(537,590){\line(-5, 3){ 16.912}}
\put(460,640){\line(-5,-3){ 16.912}}
\put(443,630){\line( 0,-1){ 20}}
\put(443,610){\line( 5,-3){ 16.912}}
\put(460,600){\line( 5, 3){ 16.912}}
\put(477,610){\line( 0, 1){ 20}}
\put(477,630){\line(-5, 3){ 16.912}}
\put(340,640){\line(-5,-3){ 16.912}}
\put(323,630){\line( 0,-1){ 20}}
\put(323,610){\line( 5,-3){ 16.912}}
\put(340,600){\line( 5, 3){ 16.912}}
\put(357,610){\line( 0, 1){ 20}}
\put(357,630){\line(-5, 3){ 16.912}}
\put(340,560){\line(-5,-3){ 16.912}}
\put(323,550){\line( 0,-1){ 20}}
\put(323,530){\line( 5,-3){ 16.912}}
\put(340,520){\line( 5, 3){ 16.912}}
\put(357,530){\line( 0, 1){ 20}}
\put(357,550){\line(-5, 3){ 16.912}}
\put(340,480){\line(-5,-3){ 16.912}}
\put(323,470){\line( 0,-1){ 20}}
\put(323,450){\line( 5,-3){ 16.912}}
\put(340,440){\line( 5, 3){ 16.912}}
\put(357,450){\line( 0, 1){ 20}}
\put(357,470){\line(-5, 3){ 16.912}}
\put(280,520){\line(-5,-3){ 16.912}}
\put(263,510){\line( 0,-1){ 20}}
\put(263,490){\line( 5,-3){ 16.912}}
\put(280,480){\line( 5, 3){ 16.912}}
\put(297,490){\line( 0, 1){ 20}}
\put(297,510){\line(-5, 3){ 16.912}}
\put(280,600){\line(-5,-3){ 16.912}}
\put(263,590){\line( 0,-1){ 20}}
\put(263,570){\line( 5,-3){ 16.912}}
\put(280,560){\line( 5, 3){ 16.912}}
\put(297,570){\line( 0, 1){ 20}}
\put(297,590){\line(-5, 3){ 16.912}}
\put(400,680){\line(-5,-3){ 16.912}}
\put(383,670){\line( 0,-1){ 20}}
\put(383,650){\line( 5,-3){ 16.912}}
\put(400,640){\line( 5, 3){ 16.912}}
\put(417,650){\line( 0, 1){ 20}}
\put(417,670){\line(-5, 3){ 16.912}}
\put(415,510){\makebox(0.4444,0.6667){\tenrm .}}
\put(415,510){\makebox(0.4444,0.6667){\tenrm .}}
\put(415,510){\makebox(0.4444,0.6667){\tenrm .}}
\put(415,510){\makebox(0.4444,0.6667){\tenrm .}}
\put(400,520){\line(-5,-3){ 16.912}}
\put(383,510){\line( 0,-1){ 20}}
\put(383,490){\line( 5,-3){ 16.912}}
\put(400,480){\line( 5, 3){ 16.912}}
\put(417,490){\line( 0, 1){ 20}}
\put(417,510){\line(-5, 3){ 16.912}}
\put(400,600){\line(-5,-3){ 16.912}}
\put(383,590){\line( 0,-1){ 20}}
\put(383,570){\line( 5,-3){ 16.912}}
\put(400,560){\line( 5, 3){ 16.912}}
\put(417,570){\line( 0, 1){ 20}}
\put(417,590){\line(-5, 3){ 16.912}}
\multiput(385,590)(-7.50000,-5.00000){7}{\makebox(0.4444,0.6667){\tenrm .}}
\multiput(325,550)(-7.50000,-5.00000){7}{\makebox(0.4444,0.6667){\tenrm .}}
\multiput(445,630)(-7.50000,-5.00000){7}{\makebox(0.4444,0.6667){\tenrm .}}
\multiput(415,590)(7.50000,-5.00000){7}{\makebox(0.4444,0.6667){\tenrm .}}
\multiput(355,630)(7.50000,-5.00000){7}{\makebox(0.4444,0.6667){\tenrm .}}
\multiput(475,550)(7.50000,-5.00000){7}{\makebox(0.4444,0.6667){\tenrm .}}
\multiput(505,590)(-7.50000,-5.00000){7}{\makebox(0.4444,0.6667){\tenrm .}}
\multiput(445,550)(-7.50000,-5.00000){7}{\makebox(0.4444,0.6667){\tenrm .}}
\multiput(385,510)(-7.50000,-5.00000){7}{\makebox(0.4444,0.6667){\tenrm .}}
\multiput(295,590)(7.50000,-5.00000){7}{\makebox(0.4444,0.6667){\tenrm .}}
\multiput(355,550)(7.50000,-5.00000){7}{\makebox(0.4444,0.6667){\tenrm .}}
\multiput(415,510)(7.50000,-5.00000){7}{\makebox(0.4444,0.6667){\tenrm .}}
\multiput(505,510)(-7.50000,-5.00000){7}{\makebox(0.4444,0.6667){\tenrm .}}
\multiput(445,470)(-7.50000,-5.00000){7}{\makebox(0.4444,0.6667){\tenrm .}}
\multiput(355,470)(7.50000,-5.00000){7}{\makebox(0.4444,0.6667){\tenrm .}}
\multiput(295,510)(7.50000,-5.00000){7}{\makebox(0.4444,0.6667){\tenrm .}}
\put(220,560){\line(-5,-3){ 16.912}}
\put(203,550){\line( 0,-1){ 20}}
\put(203,530){\line( 5,-3){ 16.912}}
\put(220,520){\line( 5, 3){ 16.912}}
\put(237,530){\line( 0, 1){ 20}}
\put(237,550){\line(-5, 3){ 16.912}}
\put(580,560){\line(-5,-3){ 16.912}}
\put(563,550){\line( 0,-1){ 20}}
\put(563,530){\line( 5,-3){ 16.912}}
\put(580,520){\line( 5, 3){ 16.912}}
\put(597,530){\line( 0, 1){ 20}}
\put(597,550){\line(-5, 3){ 16.912}}
\multiput(235,550)(7.50000,-5.00000){7}{\makebox(0.4444,0.6667){\tenrm .}}
\multiput(565,550)(-7.50000,-5.00000){7}{\makebox(0.4444,0.6667){\tenrm .}}
\put(280,680){\line(-5,-3){ 16.912}}
\put(263,670){\line( 0,-1){ 20}}
\put(263,650){\line( 5,-3){ 16.912}}
\put(280,640){\line( 5, 3){ 16.912}}
\put(297,650){\line( 0, 1){ 20}}
\put(297,670){\line(-5, 3){ 16.912}}
\put(520,680){\line(-5,-3){ 16.912}}
\put(503,670){\line( 0,-1){ 20}}
\put(503,650){\line( 5,-3){ 16.912}}
\put(520,640){\line( 5, 3){ 16.912}}
\put(537,650){\line( 0, 1){ 20}}
\put(537,670){\line(-5, 3){ 16.912}}
\put(220,640){\line(-5,-3){ 16.912}}
\put(203,630){\line( 0,-1){ 20}}
\put(203,610){\line( 5,-3){ 16.912}}
\put(220,600){\line( 5, 3){ 16.912}}
\put(237,610){\line( 0, 1){ 20}}
\put(237,630){\line(-5, 3){ 16.912}}
\put(580,640){\line(-5,-3){ 16.912}}
\put(563,630){\line( 0,-1){ 20}}
\put(563,610){\line( 5,-3){ 16.912}}
\put(580,600){\line( 5, 3){ 16.912}}
\put(597,610){\line( 0, 1){ 20}}
\put(597,630){\line(-5, 3){ 16.912}}
\put(340,720){\line(-5,-3){ 16.912}}
\put(323,710){\line( 0,-1){ 20}}
\put(323,690){\line( 5,-3){ 16.912}}
\put(340,680){\line( 5, 3){ 16.912}}
\put(357,690){\line( 0, 1){ 20}}
\put(357,710){\line(-5, 3){ 16.912}}
\put(460,720){\line(-5,-3){ 16.912}}
\put(443,710){\line( 0,-1){ 20}}
\put(443,690){\line( 5,-3){ 16.912}}
\put(460,680){\line( 5, 3){ 16.912}}
\put(477,690){\line( 0, 1){ 20}}
\put(477,710){\line(-5, 3){ 16.912}}
\multiput(235,630)(7.50000,-5.00000){7}{\makebox(0.4444,0.6667){\tenrm .}}
\multiput(295,670)(7.50000,-5.00000){7}{\makebox(0.4444,0.6667){\tenrm .}}
\multiput(355,710)(7.50000,-5.00000){7}{\makebox(0.4444,0.6667){\tenrm .}}
\multiput(415,670)(7.50000,-5.00000){7}{\makebox(0.4444,0.6667){\tenrm .}}
\multiput(475,630)(7.50000,-5.00000){7}{\makebox(0.4444,0.6667){\tenrm .}}
\multiput(535,590)(7.50000,-5.00000){7}{\makebox(0.4444,0.6667){\tenrm .}}
\multiput(565,630)(-7.50000,-5.00000){7}{\makebox(0.4444,0.6667){\tenrm .}}
\multiput(505,670)(-7.50000,-5.00000){7}{\makebox(0.4444,0.6667){\tenrm .}}
\multiput(445,710)(-7.50000,-5.00000){7}{\makebox(0.4444,0.6667){\tenrm .}}
\multiput(385,670)(-7.50000,-5.00000){7}{\makebox(0.4444,0.6667){\tenrm .}}
\multiput(325,630)(-7.50000,-5.00000){7}{\makebox(0.4444,0.6667){\tenrm .}}
\multiput(265,590)(-7.50000,-5.00000){7}{\makebox(0.4444,0.6667){\tenrm .}}
\multiput(400,680)(0.00000,-9.03226){32}{\makebox(0.4444,0.6667){\tenrm .}}
\multiput(340,720)(0.00000,-9.03226){32}{\makebox(0.4444,0.6667){\tenrm .}}
\multiput(280,680)(0.00000,-9.09091){23}{\makebox(0.4444,0.6667){\tenrm .}}
\multiput(220,640)(0.00000,-9.23077){14}{\makebox(0.4444,0.6667){\tenrm .}}
\multiput(460,720)(0.00000,-9.03226){32}{\makebox(0.4444,0.6667){\tenrm .}}
\multiput(520,680)(0.00000,-9.09091){23}{\makebox(0.4444,0.6667){\tenrm .}}
\multiput(580,640)(0.00000,-9.23077){14}{\makebox(0.4444,0.6667){\tenrm .}}
\put(435,580){\makebox(0,0)[lb]{\raisebox{0pt}[0pt][0pt]{\twlrm v}}}
\put(355,580){\makebox(0,0)[lb]{\raisebox{0pt}[0pt][0pt]{\twlrm v}}}
\put(420,545){\makebox(0,0)[lb]{\raisebox{0pt}[0pt][0pt]{\twlrm v}}}
\put(365,545){\makebox(0,0)[lb]{\raisebox{0pt}[0pt][0pt]{\twlrm v}}}
\put(285,365){\makebox(0,0)[lb]{\raisebox{0pt}[0pt][0pt]{\twlrm }}}
\end{picture}
\bigskip

Clearly what we obtain is nothing but the levels given by the
 character formula  where the singular vectors exist.
Each vertex  of the hexagons
 on the diagram  denotes a singular vector which is
embedded into the top module.
(For a much more detailed analysis see our
forthcoming paper \cite{Me}).
Since the singular vectors are unique at a certain level
we constructed all the singular vectors given by the
Kac determinant.

Now we turn to the analysis of the case  when  $W_0$ is
not diagonalisable among the singular vectors. The
problem arises only when we have two singular vectors at the same
level. This can be described by the weight $(a,a)$ and it
has $w=0$. We remark that if $w=0$ in $\v{hw}$ and we have a
singular vector at a certain level then we have another
one on the same level changing every $W$ mode to its negative. This
shows that the eigenvalues of the singular
vectors are inverse of each other. Without loss of generality
we consider  the case when the singular vectors are at an even
level. We split the eigenvectors into even and odd part
as: $(P_++xP_-)\v{hw}$ and $x$ is defined such a way that
\e
W_0 P_-\v{hw}=P_+\v{hw} \ ;\quad W_0 P_+\v{hw}=\gamma P_-\v{hw}
\label{par}
\ee
Since this state is an eigenvector of $W_0$, with eigenvalue $x$,
 we have $\gamma =x^2$.
There is another singular vector namely the  $(P_+-xP_-)\v{hw}$
state with eigenvalue $-x$. The $W_0$ eigenvalues of the singular
vectors are $\a (a-t)a$ and $-\a (a-t)a$, respectively.
If the eigenvectors correspond to the same eigenvalues then
$x = 0$ and necessarily $t=a$. In this case the only h.w.\
singular vector is $P_+$ because
the h.w.\  singular vector at level $a$ always has $(W_{-1})^a$
leading term.   This shows that $xP_- \to 0$ when $x \to 0$.
 However from  this it follows that $P_-$ is finite since
if $P_-$ were not
 finite then necessarily
 $x^{\beta}P_-$, $0<\beta <1$,
 would be a finite non zero eigenvector of $W_0$ with
zero eigenvalue (\ref{par}), but this is excluded since the only eigenvector
of $W_0$ with zero eigenvalue is $P_+$. Since $P_-$ is obviously not zero
(\ref{par}), it is a singular vector.
 This singular vector can be obtained in the $x \to 0$
$t \to a$ limit from the singular vectors defined earlier and
$W_0$ maps it into the h.w.\  singular vector. This shows that they
are in the same Jordan cell. The case when the singular vectors are
at odd level is the same except for changing even with
odd.

In summary, we consistently established a framework for the \WA
  containing complex powers of one of the generators. In this
extended space we succeeded in generating singular vectors given by
the Kac formula. The generalisation of the method for other $W$
algebras is in progress \cite{Me}.

I would like to thank Gerard Watts, Horst Kausch, Adrian Kent and
Koos de Vos for stimulating discussions and useful comments and
especially Peter Goddard for the hospitality in Cambridge.
This work was supported by
the Sz\'echenyi Istv\'an \"Oszt\"ond\'{\i}j Alap\'{\i}tv\'any.

\end{document}